\documentclass[twoside]{article}
\usepackage{Proc_NTSE_16}
\usepackage{color}
\begin{document}
\thispagestyle{plain}
%Please use the command \publref{myfilename} to print the reference to your proceedings contribution 
%at the bottom of the page where myfilename should be replaced by the name of your LaTeX file
%(e.g., use the command \publref{Johns} if the LaTeX file of your contribution called Johns.tex):
\publref{Vary}

\begin{center}
{\Large \bf \strut
%Insert the title of your contribution here
{\itshape Ab Initio} No Core Shell Model\\ with Leadership-Class Supercomputers\label{JPV}
 \strut}\\
\vspace{10mm}
{\large \bf 
%Insert the authors here. Use upper indexes a, b, c, etc., to bind authors with their addresses
% as shown below.
James P. Vary$^{a}$, Robert Basili$^{a}$, Weijie Du$^{a}$, \strut\\
Matthew Lockner$^{a}$, Pieter Maris$^{a}$, Dossay Oryspayev$^b$, Soham Pal$^{a}$, Shiplu Sarker$^{a}$,
Hasan~Metin~Aktulga{$^c$},
\strut Esmond~Ng$^d$, Meiyue Shao$^d$, Chao~Yang$^d$ 
}
\end{center}

\noindent{% Insert the addresses  here.
\small $^a$\it Dept. of Physics and Astronomy, Iowa State University, Ames, IA   50011,  USA} \\
{\small $^b$\it Dept. of Computer Science, Iowa State University, Ames, IA, 50011,  USA}\\
{\small $^c$\it Dept. of Computer Science, Michigan State University, East Lansing, MI 48824, USA} \\
{\small $^d$\it Lawrence Berkeley National Laboratory, Berkeley, CA 94720, USA}\\

%The next command defines running titles:
\markboth{
%Put here the list of authors that will be displayed in running titles:
James P. Vary {\it et al.}}
{%Put here the short title of your contribution that will be displayed in running titles:
{\it Ab Initio} No Core Shell Model} 

\begin{abstract}
Nuclear structure and reaction theory is undergoing a major renaissance with advances in many-body methods, strong interactions with greatly improved links to Quantum Chromodynamics (QCD), the advent of high performance computing, and improved computational algorithms.  Predictive power, with well-quantified uncertainty, is emerging from non-perturbative approaches along with the potential for guiding experiments to new discoveries.  We present an overview of some of our recent developments and discuss challenges that lie ahead.  Our foci include: (1) strong interactions derived from chiral effective field theory; (2) advances in solving the large sparse matrix eigenvalue problem on leadership-class supercomputers; (3) selected observables in light nuclei with the JISP16 interaction; (4) effective electroweak operators consistent with the Hamiltonian; and, (5) discussion of $A=48$ system as an opportunity for the no-core approach with the reintroduction of the core.  
\\[\baselineskip] 
\end{abstract}

\section{Introduction}
With continuing advances in Leadership-Class Supercomputers and plans for further developments leading to Exascale systems (defined as having capabilities for $10^{18}$ 
floating-point operations per second (flops)), theoreticians are developing quantum many-body approaches that portend a new era of research and discovery in physics as well as in other disciplines. In particular, the nuclear physics quantum many-body problem presents unique challenges that include the need to simultaneously develop (1) strong inter-nucleon interactions with ties to QCD in order to control the concomitant freedoms; (2) non-perturbative methods that respect all the underlying symmetries such as translational invariance; and (3) new algorithms that prove efficient in solving the quantum many-body problem on Leadership-Class Supercomputers.  This triad of forefront requirements impels multi-disciplinary collaborations that include physicists, applied mathematicians and computer scientists.

While the physics goals for computational nuclear structure and reactions may seem obvious --- i.e., retaining predictive power and quantifying the uncertainties, the opportunities and challenges presented with the continuing rapid development of supercomputer architectures is less obvious to the broader community so we will introduce some of these issues in this work. 
With the need to develop and apply fully microscopic approaches to heavier nuclei as well as to include multi-nucleon interactions and coupling to the continuum, even Exascale computers will be insufficient to meet all our plans. We therefore must also work to develop renormalization schemes that reduce the computational burden without loss of fidelity to the underlying theory.

\section{Strong Inter-Nucleon Interactions Linked to QCD}

Major theoretical advances have been made in the last few years in developing the theory of nuclear strong interaction Hamiltonians from the underlying theory QCD using chiral effective field theory (EFT)  \cite{EpHa09,MaEn11}. Chiral EFT provides a hierarchy of two-nucleon ($NN$), three-nucleon ($3N$), four-nucleon ($4N$) interactions, etc., with increasing chiral order where chiral order is defined in terms of a dimensionless parameter $Q/\Lambda$.  Here $Q$ represents a characteristic low-momentum scale, which is frequently taken to be the mass of the pion or the momentum transfer in the case of scattering, and $\Lambda$ is the confinement (symmetry breaking) scale of QCD which is usually in the range of 4--7 times the mass of the pion. Most recently, a new generation of chiral interactions is becoming available 
\cite{Maris:2016wrd,Binder:2015mbz} that aims for improved consistency of the $NN$ and multi-$N$ interactions. These developments motivate us to adopt chiral EFT Hamiltonians in our current and planned applications.

\begin{figure}[!t]
\centerline{\includegraphics[width=0.94\textwidth,angle=90]{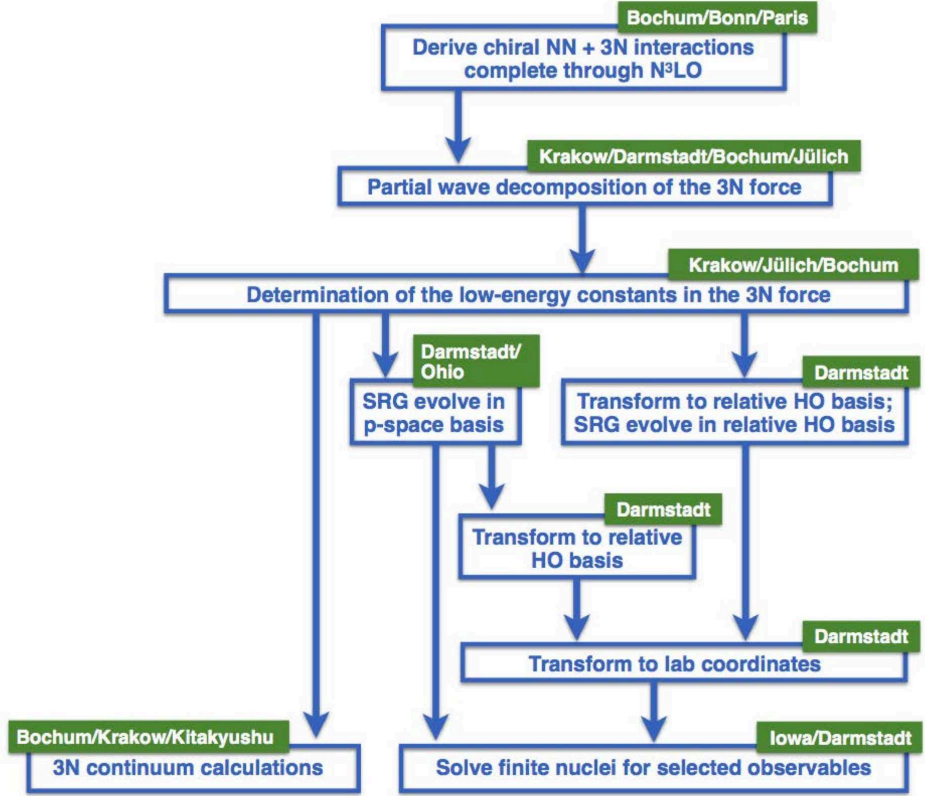}}
\caption{Workflow of the Low Energy Nuclear Physics International Collaboration (LENPIC) depicting a one-pass effort \cite{LENPIC}. Multiple passes through the entire workflow and/or subareas of the workflow are performed in order to arrive at a final regulated chiral EFT interaction with quantified uncertainties in the LECs.}
\label{LENPIC_work}      
\end{figure}

One hallmark of the development of the newest generation of chiral Hamiltonians is the close collaboration of the few-body teams traditionally leading the Hamiltonian developments and the many-body applications teams that have traditionally been on the receiving end of the Hamiltonians once they are released.  This teamwork is exemplified by the Low Energy Nuclear Physics
International Collaboration (LENPIC)~\cite{Binder:2015mbz,LENPIC} which has a workflow portrayed in Fig. \ref{LENPIC_work}.
In this new paradigm, there is a close interplay between the Hamiltonian developers and the many-nucleon applicators so that there is now feedback on important issues such as the choice of regulators and the determination of the Low-Energy Constants (LECs) that cannot yet be determined directly from QCD.  In principle, this will lead to a selection of the ingredients in the chiral EFT that are more harmonious with improved convergence rates, predictive power and quantified uncertainties.

At the present time, only the new chiral $NN$ interactions are available~\cite{Epelbaum:2014sza} and the consistent chiral $3N$ and $4N$ interactions are under development with an expected release in 2018. The results with the new chiral $NN$ interactions 
are very encouraging
yet still indicate the need for consistent $3N$ interactions to accurately describe the properties of light nuclei 
\cite{Maris:2016wrd,Binder:2015mbz}. In order to reach such a conclusion, new methods of uncertainty quantification were 
developed and applied~\cite{Epelbaum:2014sza,Maris:2016wrd,Binder:2015mbz}.
For the purposes of this work we will adopt alternative state-of-the-art $NN$ interactions to illustrate calculated nuclear properties and uncertainty quantification with Leadership-Class Supercomputers.

\section{{\it ab initio} No Core Shell Model}

The {\it ab initio} No Core Shell Model (NCSM) formulates the nuclear quantum many-body problem as a non-relativistic Hamiltonian eigenvalue problem in an adopted basis space (most frequently a harmonic oscillator (HO) basis) where all nucleons in the nucleus are treated on the same footing \cite{Navratil:2000ww,Navratil:2000gs,Navratil:2007we,Maris:2008ax,Maris:2009bx,Barrett:2013nh,Maris:2013poa,Shirokov_review_JPV382:2014}. This representation of the Hamiltonian in a basis, using $NN$, $3N$ and $4N$ interactions, generates a large sparse matrix eigenvalue problem for which we seek the low-lying eigenvalues and eigenvectors in order to compare with experimental data and to make testable predictions. 

Since the interactions are strong, inducing short-range correlations, the challenge is to perform the calculations in a sufficiently large basis to obtain convergence.  Alternatively, one may perform a sequence of calculations in ever-increasing basis spaces and extrapolate the eigenvalues, as well as other observables, to the infinite matrix limit. We refer to this approach for obtaining the converged results and quantified uncertainties as the No-Core Full Configuration (NCFC) method. 

The reach of the NCFC method with fixed uncertainty is limited by the available computational resources.  To minimize uncertainties while increasing the range of accessible atomic numbers $A$, we seek to efficiently use the largest and fastest available supercomputers.

\begin{figure}[!h]
\centering
\rotatebox{90}{\parbox{1.\textheight}{\centerline{
\parbox{0.99\textheight}{\includegraphics[clip,width=0.98\textheight]{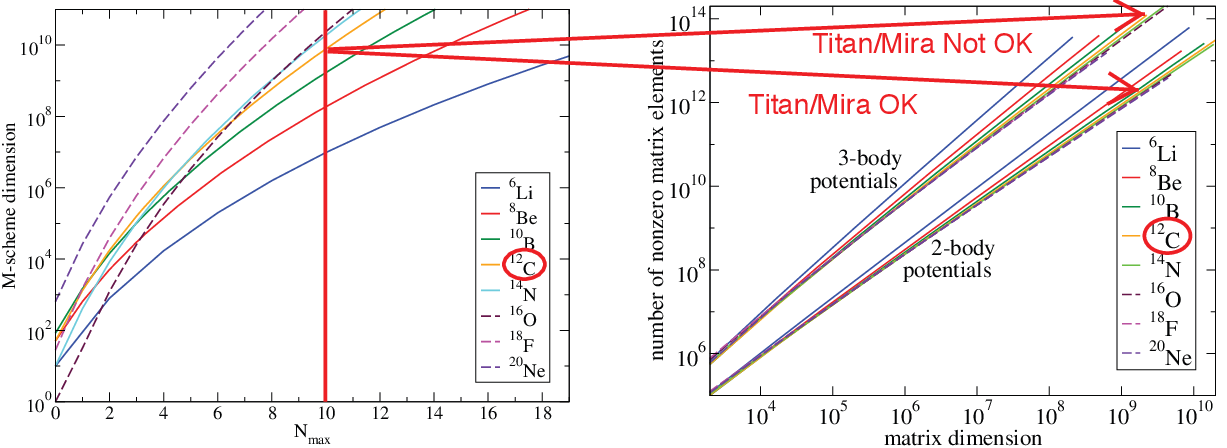}}}
%\parbox{0.99\textheight}{\includegraphics[clip,width=0.98\textheight]{NCSM_Scaling_Props_v2.eps}}}
\caption{{\it Left panel} presents the Hamiltonian matrix dimension for a basis with total 
angular momentum projection $M=0$ as a function of $N_{\rm max}$ for a selection of
even-$A$ nuclei. {\it Right panel} displays the number of non-zero matrix elements as 
a function of $M=0$ matrix dimension for the same cases as the left panel with either 
$NN$-only (``2-body potentials'') or $3N$ interactions (``3-body potentials''). The arrows
stretching from the {\it left panel} to the {\it right panel} indicate the supercomputers
on which that case will not fit (``Not OK'') or will fit (``OK'') within available memory.  
Mira is located at Argonne National
Laboratory and Titan is located at Oak Ridge National Laboratory.}
\label{NCSM_Scaling_Props}}}      
\end{figure}
To achieve this goal within a constantly evolving Leadership-Class Supercomputing environment (see following section) 
requires collaborations of physicists, computer scientists and applied mathematicians.  Such collaborations have resulted 
in a string of successes in the areas of eigensolver algorithms, memory management and communications \cite{Sternberg:2008,Vary:2009qp,Maris_JPV:2010,Aktulga_JPV324:2011,Maris:2012du,Aktulga_JPV337:2012,Maris:2013a,
Oryspayev_JPV358:2013,Aktulga_JPV366:2013,Oryspayev:2015xxx,
Potter:2014gwa,
Potter:2014dwa,Yang_JPV379:2014,
Oryspayev_JPV402:2015,
Vary:2015dda,
Aktulga_JPV432:2016,Shao_JPV443:2017}.

In order to characterize the level of effort required to achieve a target level of uncertainty, we can take the example of the NCSM/NCFC application to light nuclei within a HO basis where we employ a many-body cutoff parameter 
$N_{\rm max}$.  $N_{\rm max}$ is defined as the maximum number of HO quanta (summed over the single-particle states in each basis state) allowed above the minimum needed to satisfy the Pauli principle.  The basis is also constrained by total parity and total angular momentum projection $M$.  The latter constraint is available since we work in an $M$-scheme basis rather than a basis of good total angular momentum $J$.  With a given choice of $M$, all states of good $J \geq |M|$ are accessible in the same calculation and we evaluate $J$ in a post-analysis using the produced eigenfunctions. Once $N_{\rm max}$ and the other constraints are determined, the matrix dimension is known.  The {\it left panel} of Fig. \ref{NCSM_Scaling_Props} shows a semi-log plot of the rapid rise of matrix dimension with $N_{\rm max}$ at $M=0$ for natural parity states in a selection of nuclei.  In order to obtain
convergence for bound states with realistic interactions (those that accurately describe $NN$ scattering) and achieve a reasonable uncertainty, we find it highly desirable to have results at $N_{\rm max}=10$ or above as indicated by the vertical line in the {\it left panel} of Fig. \ref{NCSM_Scaling_Props}.

The {\it right panel} of Fig. \ref{NCSM_Scaling_Props} illustrates a useful measure of the computational effort --- the number of non-zero (NNZ) many-body matrix elements as a function of the matrix dimension.  Here we adopt the same cases shown in the {\it left panel} and present the NNZs for both $NN$-only calculations and calculations with $3N$ interactions.  Note that the NNZs rise with nearly linear trajectories on this log-log plot and they are tightly bunched so as to suggest reasonable independence of $A$ for each trajectory.  Since the computational effort (consisting of both  the time to evaluate and store the many-body Hamiltonian, and the amount of memory needed) is based primarily on the NNZs, we can estimate the computational resources needed once the matrix dimension is known (as in the {\it left panel} of Fig. \ref{NCSM_Scaling_Props}) and the interaction is specified.  This process is illustrated by the arrows reaching from the {\it left panel} to the 
{\it right panel} of Fig. \ref{NCSM_Scaling_Props} for the case of $^{12}$C 
at $N_{\rm max}=10$ for either a pure $NN$ or a $3N$ interaction.  With the NNZs fixed, we know whether a given calculation fits within the memory of the chosen Leadership-Class Supercomputer as indicated by the labels on the two arrows.  With the requirement to store the many-body Hamiltonian in core and to use it for the diagonalization process on Titan or Mira, we determine that we can solve for the low-lying spectra of $^{12}$C at $N_{\rm max}=10$ with an $NN$-only interaction but not with $3N$ interactions.  A simple functional form relating the matrix dimension D to the NNZs for 2-body interactions is \cite{Vary:2009qp}
\begin{equation}
\mathrm{NNZ} =D+D^{1+\frac{12}{14+\ln{D}}},
\label{NNZscaling}
\end{equation}
where $D$ is the matrix dimension.

It should be noted that these NCSM/NCFC successes in low-energy nuclear physics have applications in other areas of strong-interaction physics.  For example, Hamiltonian methods are gaining popularity in non-perturbative solutions of quantum field theory \cite{Vary:2008zz,Vary:2009zz,Vary:2009gt,Vary:2009qz,Vary:2010zza,Honkanen:2010rc,Zhao:2011ct,Maris:2013qma,Vary:2014tqa,Zhao:2014xaa,Chakrabarti:2014cwa} motivated, in part, by the advances being made by our teams in solving the {\it ab initio} NCSM/NCFC. Recent applications, in what is called the  Basis Light-Front Quantization (BLFQ) approach \cite{Vary:2008zz,Vary:2009zz,Vary:2009gt}, include non-perturbative solutions of positronium at strong coupling \cite{Maris:2013qma,Li:2013cga,Wiecki:2013cba,Wiecki:2015xxa,Wiecki:2014ola,Adhikari:2016idg} and solutions for the mass spectra, decay constants, form factors and vector meson production rates for heavy quarkonia \cite{Li:2015zda,Vary:2016emi,Vary:2016ccz,Chen:2016dlk,Chen:2017mat,Li:2016wwu,Li:2017mlw,Leitao:2017esb}. Remarkably, results for QED in the BLFQ approach have been achieved with Hamiltonian matrix dimensions exceeding $18$ billion basis states \cite{Zhao:2014xaa,Chakrabarti:2014cwa}. 

In addition to the use of relativistic Hamiltonian methods for static properties of strongly-interacting systems, time-dependent scattering with strong fields in quantum field theory has been introduced and successfully applied using the interaction picture. This is referred to as the time-dependent BLFQ (tBLFQ) approach \cite{Zhao:2013cma,Vary:2013kma,Zhao:2013jia,Chen:2017uuq}.  In the tBLFQ approach, one first solves the relevant bound state problems in BLFQ and then evolves the system in light-front time with the possible addition of strong time-dependent external fields.  This quantum time evolution approach leads to the total scattering amplitude from which projections to specific final channels can be performed and relativistic observables evaluated. Analogous development and applications of a time-dependent NCSM approach to non-relativistic strong interaction problems is underway \cite{Du_etal_here} adapting techniques from tBLFQ.

Following the next two sections devoted to a perspective on supercomputer resources (Sec. 4) and algorithm improvements (Sec. 5), we present a selection of recent results 
and outline challenges that lie ahead.  
Our aim with this limited choice of applications is to complement other presentations at this meeting that cover closely-related topics.  We note especially the papers at this meeting related to the NCSM/NCFC, new Hamiltonians and NCSM extensions to scattering theory
by Shirokov~\cite{Shirokov_here}, 
by Skibinski~\cite{Skibinski_here}, 
by Zhao~\cite{Du_etal_here},
by A. Mazur~\cite{A.Mazur_here},
by I. Mazur~\cite{I.Mazur_here},
and by Kulikov~\cite{Kulikov_here}.  
We therefore focus here on the following recent results: 
(1) nuclear binding energies, excitation energies and magnetic moments of light nuclei
with a realistic $NN$ interaction;
(2) construction of effective electroweak interactions for nuclear moments
and transitions; and
(3) outline of an approach for calculating $A=48$ nuclei for evaluating nuclear double beta-decay matrix elements both with and without neutrinos.

\section{Leadership-Class Supercomputers}
The list of the world's top 500 supercomputers is updated every six months\cite{Super500} where one observes that China's
TaihuLight has topped the list for the past few cycles.  TaihuLight has more than 10 million cores and is rated at 93 PetaFlops or 
nearly $10^{17}$ floating point operations per second.  In the United States we currently refer to Leadership-Class Supercomputers as those rated at about one-tenth of the TaihuLight rating.  For the United States, this includes facilities available for general scientific computing such as Titan at Oak Ridge National Laboratory (rated number~4 with 17.6 PetaFlops),
Cori at Lawrence Berkeley National Laboratory/NERSC (rated number~6 with 14 PetaFlops) and Mira at Argonne National Laboratory (rated number~9 with 8.6 PetaFlops).  Researchers using other Leadership-Class Supercomputers are also attending this meeting and will likely credit their own facilities while the results that we present have most frequently been produced on these three above-mentioned US facilities by our group at Iowa State University and by our collaborators.

\begin{figure}[!t]
\centerline{\includegraphics[width=0.94\textwidth]{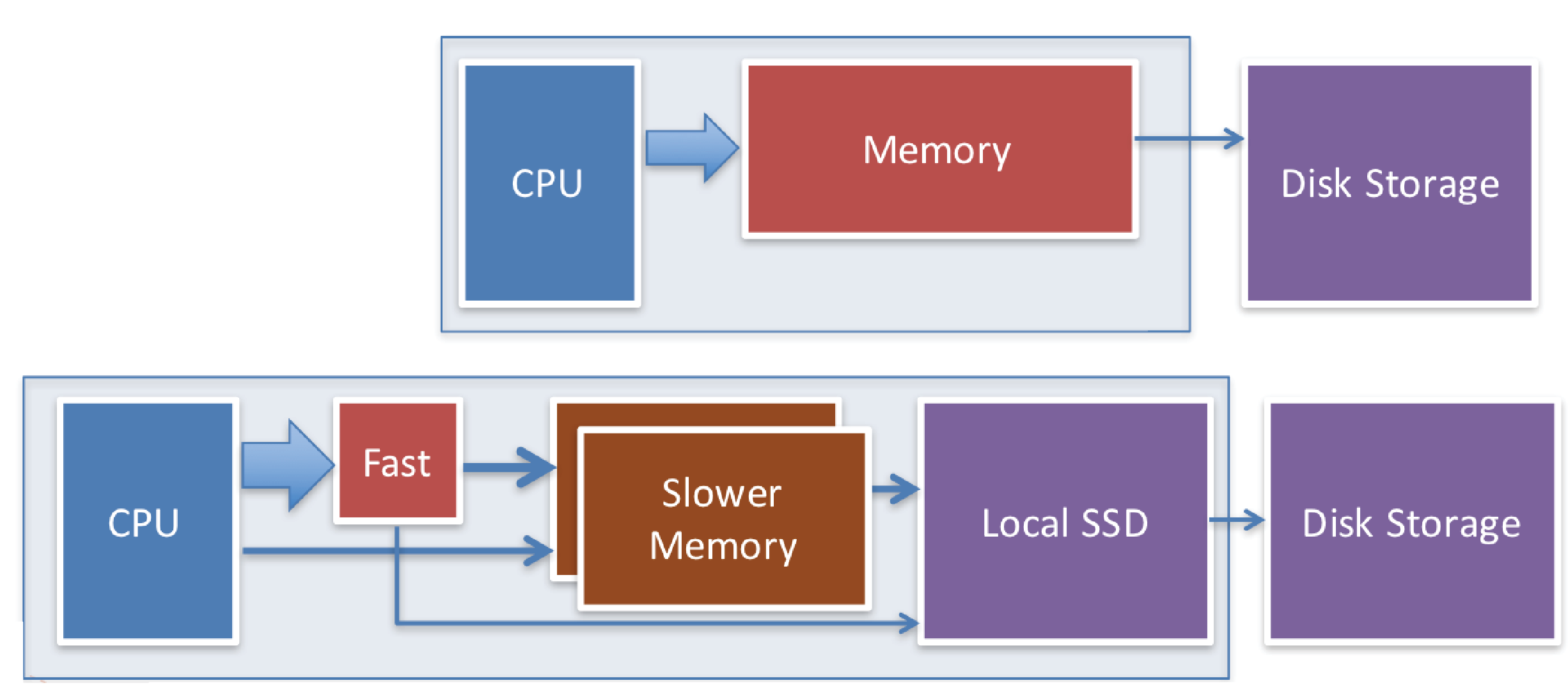}}
\caption{Sketch of the change in computer architecture providing new levels of challenges
for algorithms and software.  The traditional single-level of random-access memory (RAM) is
replaced by a memory hierarchy that, to be used efficiently, requires careful analysis of 
data locality and usage intensity.}
\label{Architecture}      
\end{figure}

Here, we would also like to mention that each facility has a different architecture and that each architecture requires extensive efforts by physicists, applied mathematicians and computer scientists to enable forefront research with efficient algorithms and finely-tuned parallel computing codes.  For these purposes, we have benefitted greatly from more than ten years of support from the US Department of Energy's SciDAC program \cite{SciDAC_site} that supports the collaborative research on the {\it ab initio} NCSM/NCFC algorithms and codes keeping them competitive over cycles in disruptive architecture changes. As an illustration of some of the newer architectures, Fig. \ref{Architecture} sketches the move into hierarchical memories.  Multiple communication topologies within nodes and among nodes further increase the complexities.

While today's Leadership-Class Supercomputers are certainly impressive technological achievements empowering forefront discoveries, there is a race to design, fund and build even larger machines to reach the Exascale capability level of $10^{18}$ floating point operations per second, more than an order of magnitude increase over the current top supercomputer, TaihuLight, in China.  Policies have been announced to achieve this goal within 5 years.  Past experience supports the belief that the technology will be further disruptive and will require major efforts by the same teams at work today in order to achieve forefront physics results with algorithms and codes that run efficiently at Exascale.  Those efforts have to begin years before the machine comes into operation in order to fully capitalize on the major investments to design, build and operate it.  Fortunately, the US Department of Energy is continuing its support through SciDAC and we can remain optimistic that theoretical nuclear physics will benefit greatly from the Exascale machine when it is delivered.  In the interim, Leadership-Class Supercomputers with capabilities in the hundreds of PetaFlops are under construction now and will become available in 2018--19 to provide an intermediate step from the current machines to the Exascale machines and we plan to fully utilize these new facilities for {\it ab initio} nuclear structure and nuclear reactions. 

\section{Algorithmic improvements for the NCSM/NCFC}
Efficient methods to construct and diagonalize the sparse nuclear Hamiltonian of the {\it ab initio} NCSM
on Leadership-Class Supercomputers have been implemented in the software package MFDn 
(Many Fermion Dynamics for nuclear structure) \cite{Maris:2013a,Maris:2012du,Vary:2015dda}. 
MFDn uses the Lanczos algorithm \cite{Lanczos1950,Saad2011} to compute 
the desired eigenvalues and eigenvectors. Using the eigenvectors, MFDn then produces 
additional experimental observables such as electromagnetic and weak interaction transition
rates. There is flexibility to use only $NN$ interactions or $NN$ plus 3$N$ interactions as input.

Over the last several years, we have developed a number of techniques to improve 
the computational efficiency of MFDn including:

\begin{itemize}
\item{an efficient scalable parallel scheme for constructing the Hamiltonian matrix \cite{Sternberg:2008}},
\item{efficient data distribution schemes that take into account the topology of
the interconnect \cite{Aktulga_JPV337:2012}},
\item{techniques to overlap communication with computation in a hybrid MPI/
OpenMP programming model \cite{Aktulga_JPV366:2013,Oryspayev:2015xxx}},
\item{an efficient scheme to multiply the sparse matrix Hamiltonian with a number
of vectors \cite{Aktulga_JPV432:2016}},
\item{introduction of an accelerated eigensolver that employs a preconditioned 
block iterative method \cite{Shao_JPV443:2017}}.
\end{itemize}

\begin{figure}[!t]
\includegraphics[width=0.38\textwidth]{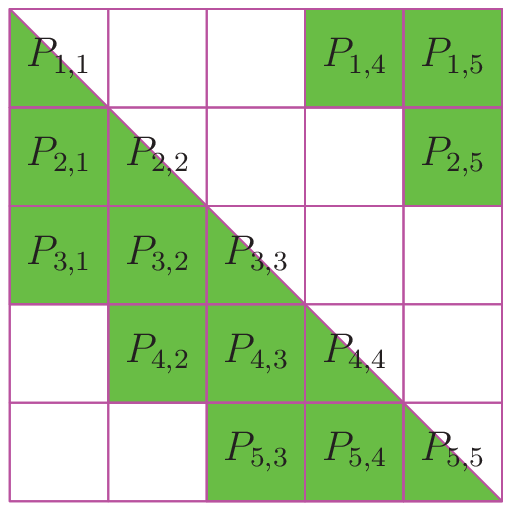}\qquad
\includegraphics[width=0.55\textwidth]{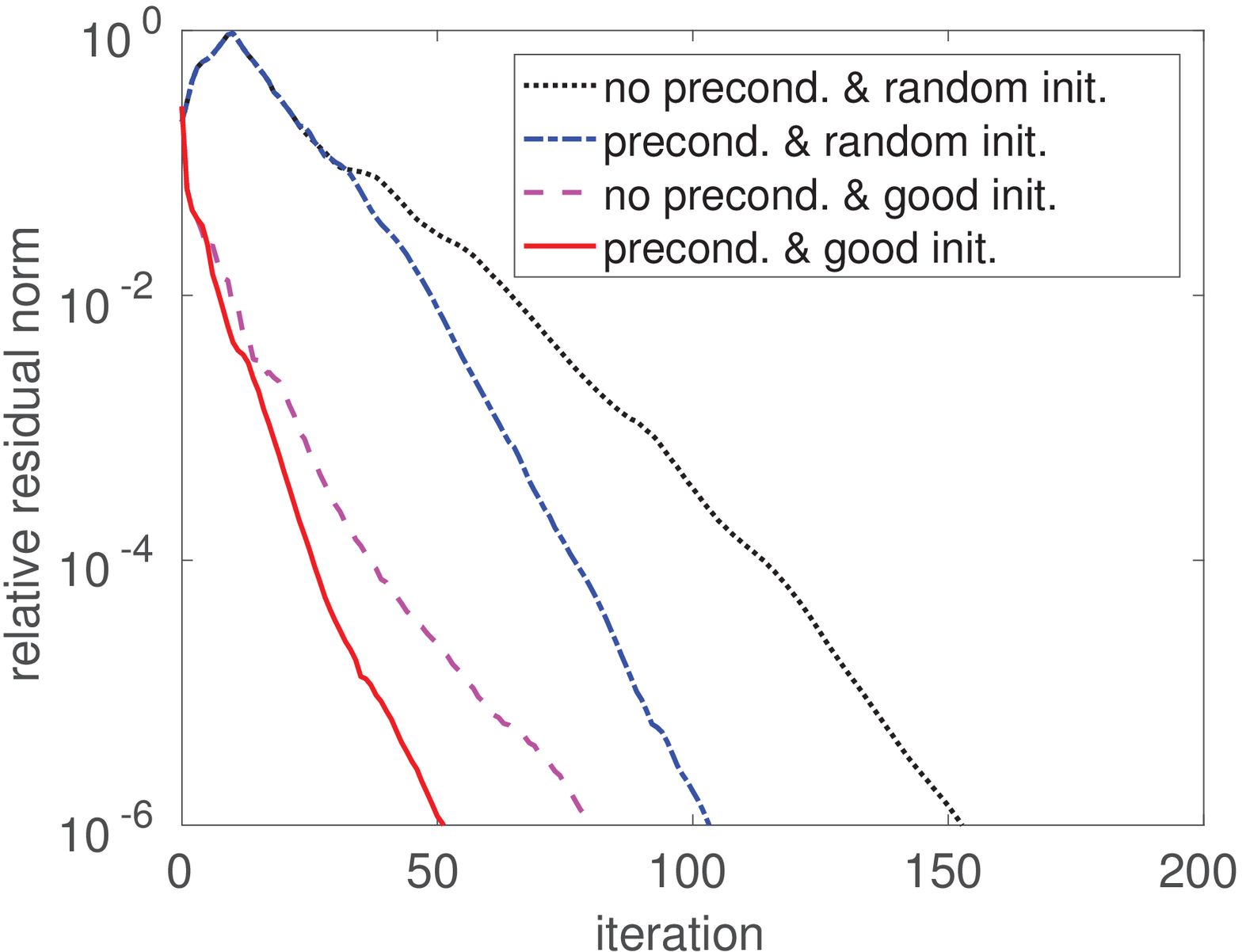}
\caption{{\it Left panel} displays the assignment of matrix elements of a symmetric
matrix (green shaded regions) among MPI ranks (squares) that achieves a balance in
communications during the Lanczos process. {\it Right panel} shows improved convergence
rates of the LOBPCG algorithm achieved by preconditioning and by good initialization
\cite{Shao_JPV443:2017}.}
\label{algorithms}      
\end{figure}

As the number of cores has been increasing dramatically during the past decade, one faces
an increasing challenge to minimize the time spent on inter-processor communications.
Among our accomplishments, we developed distribution schemes for the computations that
reduce communication times as illustrated in the {\it left panel} of Fig. \ref{algorithms} where
we sketch the distribution of unique partitions of the symmetric matrix among processors 
$P_{ij}$.  This distribution achieves a balance of the MPI reduce (and subsequent broadcast) 
operations for rows and columns of processors that perform the matrix--vector multiplies for both 
the Hamiltonian matrix and its transpose.

The most recent development \cite{Shao_JPV443:2017} introduces a new eigensolver
into MFDn, the Locally Optimal Block Preconditioned Conjugate Gradient (LOBPCG)
algorithm \cite{Knyazev2001}. The use of a block iterative method allows us to improve the memory 
access pattern of the computation and make use of approximations to several eigenvectors
at the same time. To make this algorithm efficient, as shown in the {\it right panel} of Fig. \ref{algorithms}, 
we identified an effective preconditioner coupled with techniques to generate good initial guesses 
that significantly accelerate the convergence of the LOBPCG algorithm on large-scale distributed-memory
clusters.

Further efforts are underway to speed up communications among nodes and to develop a 
post processor for efficiently evaluating transitions between nuclear systems.  The latter 
is needed for planned calculations of the nuclear matrix elements for double-beta
decay, both with and without neutrinos.  Additional efforts are underway to develop scripts 
for a broad set of standard applications that
facilitate conversion from one architecture to another.

\begin{figure}[!t]
\rotatebox{90}{\parbox{1.\textheight}{\centerline{
\parbox{0.99\textheight}{\includegraphics[clip,width=0.98\textheight]{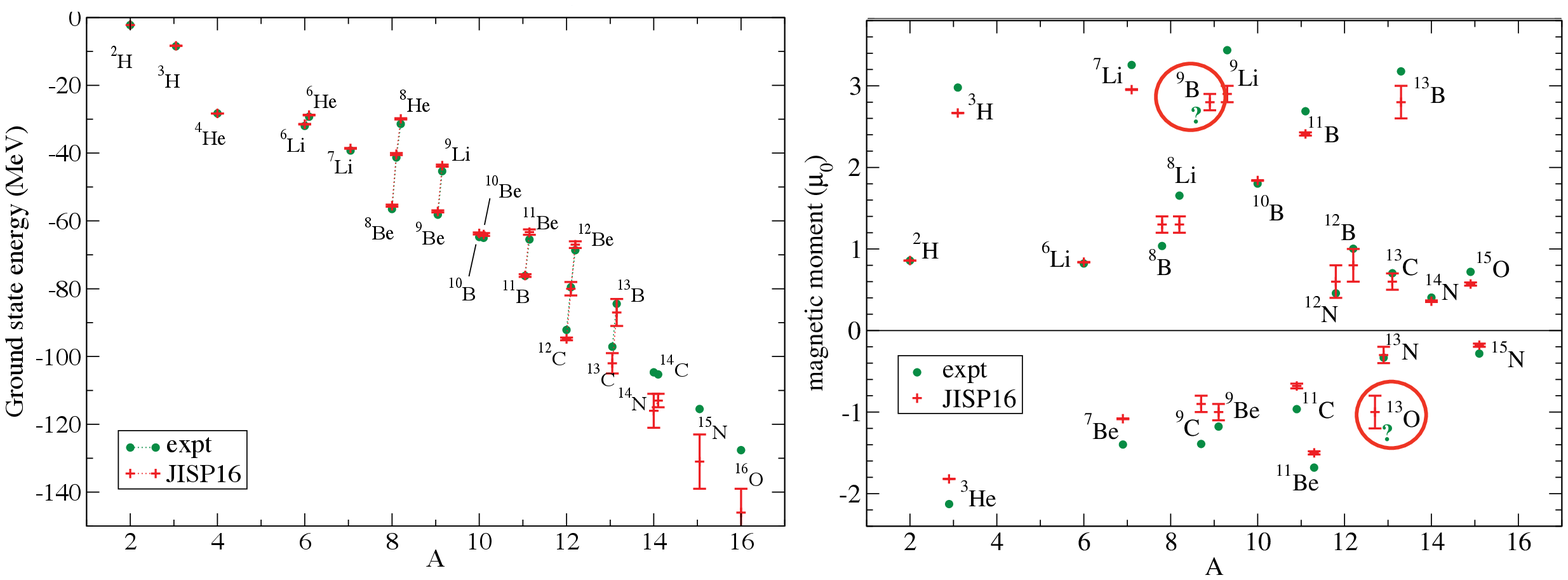}}}
\caption{NCFC results for the ground states of light nuclei obtained with the JISP16 interaction (red points) compared with experiment (green points) where available~\cite{Maris:2013poa}. Uncertainties in NCFC extrapolations are indicated by the red error bars. {\it Left panel} displays ground state energies while the {\it right panel} shows magnetic moments. Note that two
cases are circled to indicate that they represent predictions awaiting experimental results for comparison.}
\label{Summary_JISP16_results} }}     
\end{figure}

\section{Results for light nuclei with JISP16}

In this section, we briefly review selected results for light nuclei using the realistic JISP16 $NN$ interaction \cite{Shirokov:2004ff,Shirokov:2005bk} within the NCFC approach \cite{Maris:2008ax,Maris:2009bx,Maris:2013poa,Shirokov_review_JPV382:2014}. Fig. \ref{Summary_JISP16_results} presents ground state energies for 24 light nuclei in the {\it left panel}.  While JISP16 was tuned with phase-equivalent transformations to the properties of nuclei up to $A=7$, it was only approximately tuned to the ground state energy of $^{16}$O.  It is therefore not surprising that JISP16 overbinds nuclei at the upper end of the p-shell beginning with $A=10$.  We note that the recently-developed Daejeon16 $NN$ interaction succeeds in  improving the agreement between theory and experiment for the ground state energies of the p-shell 
nuclei as well as other properties of light nuclei~\cite{Shirokov:2016ead}.

Applications of JISP16 to the Lithium isotopes and the Berylium isotopes already have an extensive track record due both to experimental interests and to NCFC advances that provide results with increasing precision over time. An earlier detailed investigation of the Lithium isotopes with JISP16 \cite{Cockrell:2012vd} provides NCFC 
results that serve as a baseline for
recent extensive investigations of $^6$Li \cite{Shin:2016poa} as well as $^7$Li and $^7$Be \cite{Heng:2016umo}. Among other improvements, these recent works achieve spectral and electroweak properties in larger model spaces than previously feasible.  That is, they provide results closer to convergence which, upon extrapolation, provide NCFC observables with diminished uncertainties.

\begin{table}

{\center
\begin{tabular}{|c|c|c||c|c|c|}
\hline
  NCFC &  $N_{\rm max}$  & Ref. & HH & $K_{\rm max}$  & Ref.  \\ 
\hline
 $-31.00(31)$  &   $12$        & \cite{Shirokov:2005bk}& $-31.46(5)$            &  $14$  &  \cite{Vaintraub:2009mm}    \\
 $-31.47(9)$    &   $16$        & \cite{Maris:2008ax}     & $-31.67(3)$            &  $12$  &  \cite{Barnea:2010zz}          \\
 $-31.49(3)$    &   $16$        & \cite{Cockrell:2012vd} &                           &        &                                             \\
 $-31.49(6)$    &   $16$        & \cite{Maris:2013poa}   &                           &        &                                             \\
 $-31.42(5)$    &   $16$        & \cite{Shirokov_review_JPV382:2014}&    &        &                                             \\
 $-31.46(3)$    &   $14$        & \cite{Shin:2016poa}    &                            &        &                                             \\
 $-31.51(3)$    &   $16$        & \cite{Shin:2016poa}    &                            &        &                                             \\
 $-31.53(2)$    &   $18$        & \cite{Shin:2016poa}    &                            &        &                                             \\
\hline
\end{tabular} }
\caption{Dependence on the many-body method, on the extrapolation method,  and 
on the model space cutoff of the theoretical ground state (gs) energy (in MeV) of $^6$Li with JISP16. 
The results are arranged vertically in chronological order from earliest to most recent.
The $N_{\rm max}$ cutoff of the NCFC method and the $K_{\rm max}$ of the 
Hyperspherical Harmonics (HH) method are not directly related except 
that both should be taken to infinity to obtain the exact result. For comparison, the
experimental $^6$Li gs energy is $-31.995$ MeV and the NCFC result with Daejeon16 
using results up through $N_{\rm max}=14$  is $-31.98(2)$ MeV~\cite{Shirokov:2016ead}. 
For completeness, we note that Ref.~\cite{Shin:2016poa} quotes an extrapolated RMS
charge radius of $2.28(3)$ fm for $^6$Li which is to be compared with the experimental result
of $2.38(3)$ fm.}
\label{6Li_observables}
\end{table}

We accumulate a sample of the results for the $^6$Li extrapolated ground state (gs) energy with the JISP16 interaction 
in Table \ref{6Li_observables}.  Results from both the NCFC and the Hyperspherical Harmonics (HH) method
are included with the reference for each result quoted.  In general, there is a consistency among these results 
with the possible exception of the earliest NCFC result extrapolating from the smallest basis space.  Another
exception may be the HH result of Ref.~\cite{Barnea:2010zz} that extrapolated results obtained with the OLS renormalization (second entry in the HH column).  It is interesting to note that the NCFC results have tended to
drift towards increased binding and towards the experimental result as results from larger basis spaces 
have become available over time.  The 
difference between the experimental and theoretical ground state (gs) energy is now at 470(20) keV.  
It is also interesting to
note that the extrapolated root-mean-square (RMS) radius is tending in the direction of the experimental
result (from below) as the use of larger basis spaces become available~\cite{Shin:2016poa}.

In the {\it right panel} of Fig. \ref{Summary_JISP16_results}, we present a comparison between theory and experiment for 23 magnetic moments of states in light nuclei, where such a comparison is feasible.  In two cases, we present predictions for comparison with possible future experiments.  We evaluate these magnetic moments using only the bare operator.  Overall, the agreement is good considering the level of the approximation for the magnetic dipole operator. In the future,
we plan to incorporate 2-body current corrections.  We anticipate that these corrections will be of the order of a few percent and will further improve the agreement between theory and experiment. We base these estimations on the results presented in Ref. \cite{Pastore:2012rp} where similar differences between theory and experiment are obtained before 2-body currents are introduced.  Those 2-body currents are found to further improve the agreement between theory and experiment.

In the previous conference in this series, we reviewed \cite{Vary:2015dda} NCFC results for the 
Berylium isotopes with JISP16 where emergent collective motion is evident in the spectra, magnetic dipole moments,
M1 transitions, quadrupole moments and E2 transitions.  Recent efforts further support and extend the claims 
of emergent collective rotational behavior in the Berylium isotopes \cite{Maris:2014jha,Caprio:2015iqa,Caprio:2015jga}.  Multiple rotational bands have been identified in the NCFC calculations for both natural and unnatural parity.  
It is interesting to note that some of the bands are not observed to terminate at the angular momentum naively 
expected from nucleons populating the p-shell orbits.  Analysis of extrapolations of the NCFC results provides 
rotational model parameters in good agreement with the corresponding parameters extracted from the experimental 
data \cite{Maris:2014jha,Caprio:2015iqa,Caprio:2015jga}.

Emergent collective motion also provides inspiration for optimized basis spaces, basis spaces that offer the promise 
of accelerating convergence \cite{Dytrych:2013cca,Dytrych:2015yxa,Dytrych:2016vjy}.  
With JISP16 we have investigated truncation schemes based on SU(3) symmetry in p-shell 
nuclei.  We have found that basis space dimensions can indeed be reduced while incurring additional computational
cost for evaluating the many-body matrix elements in the SU(3) basis.  Developments are ongoing so it will be
some time before we know definitively the net gains achievable with selected SU(3) basis spaces.  In the meantime,
the more compact SU(3) representation of eigenfunctions promotes our physical intuition and knowledge of the
nuclear underlying symmetries predicted by the {\it ab initio} NCFC.

\section{Effective electroweak interactions for the NCSM}

We now turn attention to the effects that arise when consistent effective electroweak operators are included.  By consistent, we mean that the electroweak operators are evaluated in the same formalism as the strong interactions employed in the Hamiltonian.  In the case of interactions from chiral EFT, this implies that the electroweak operators are also evaluated in chiral EFT to the same chiral order as the strong interaction.  

\begin{figure}[!t]
\centerline{\includegraphics[width=1.0\textwidth]{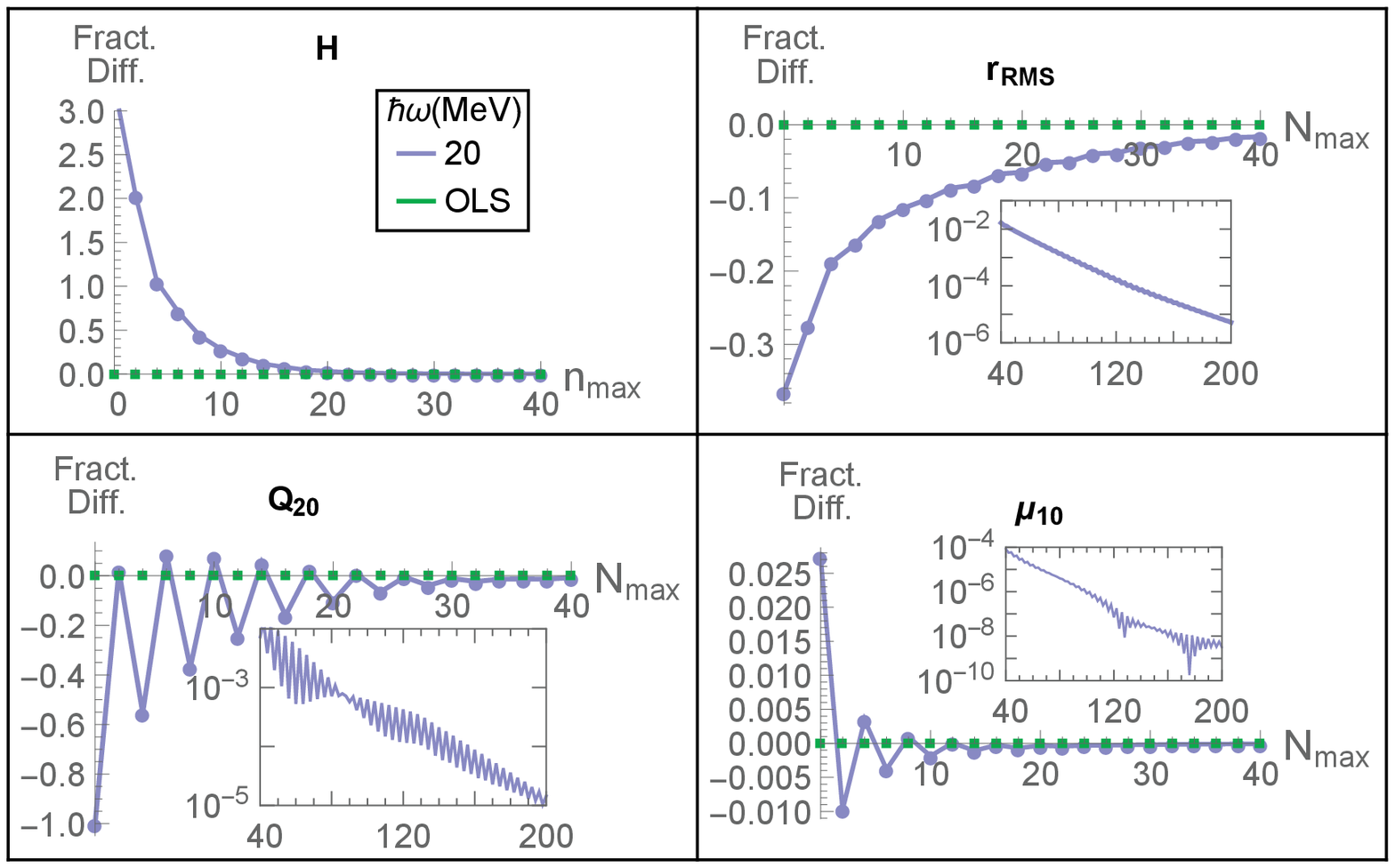}}
\caption{Comparisons of bare operators (solid dots) and OLS-renormalized operators (green squares)
for the ground state (gs) of the deuteron obtained with the LENPIC chiral N2LO interaction \cite{Binder:2015mbz}
as a function of $N_{\rm max}$ in a HO basis with $\hbar\Omega = 20$ MeV.  
These basis parameters define the model space in which the calculation is performed.  The observables
include the gs energy (upper left panel), root-mean-square radius $r_{RMS}$ (upper right panel), 
quadrupole moment  $Q_{20}$ (lower left panel) and magnetic moment $\mu_{10}$ (lower right panel).  
All results are plotted as a fractional difference "Fract.\ Diff." defined as $(model - exact)/exact$.  
We take the results at $N_{\rm max}=400$ as the exact results since they are converged to at least
8 significant digits.
The insets present the magnitude of the fractional
difference on a log scale for an extended range of $N_{\rm max}$.}
\label{Summary_OLS_effops}      
\end{figure}

Here we will provide demonstration cases using only the two-nucleon system for the present purposes.  Specifically, we study the
simple case of the gs of the deuteron solved as a matrix eigenvalue problem in the HO basis as a function of the $N_{\rm max}$ truncation.  Using the LENPIC $NN$ interaction at chiral N2LO with the regulator fixed at 1.0 fm \cite{LENPIC}, we
present the deuteron gs energy in the upper-left quadrant of Fig. \ref{Summary_OLS_effops} as a function of $N_{\rm max}$ at 
$\hbar\Omega = 20$ MeV.  As expected the gs energy converges uniformly from above with increasing $N_{\rm max}$.  In the same panel we show the gs energy results in the $N_{\rm max}$-truncated spaces following renormalization with the Okubo-Lee-Suzuki (OLS) method~\cite{Okubo:1954zz,Suzuki:1980yp,Suzuki:1982,Barrett:2013nh}. The OLS procedure produces the exact ground state energy to within numerical precision for every truncated model space.  This confirms the method is working as it should and we have numerical stability in our procedures.

Next, we apply the derived OLS transformation to additional deuteron ground state observables and display the results in the remaining three panels of Fig. \ref{Summary_OLS_effops}.  In each case, we employ only the bare operator in the present demonstration in order to gauge the size of the effects of truncation without OLS renormalization.  For $r_{RMS}$ a very small basis space results in about a 30\% reduction which slowly falls to about 1\% at about $N_{\rm max}=40$.  We stress that these results, as well as those for the other observables, are dependent on the chosen value of $\hbar\Omega$ which we have taken arbitrarily to be $20$ MeV in the present demonstration.

The quadrupole moment appears to fluctuate in the truncated model spaces which can be attributed to a sensitivity to having an odd versus an even number of $L=2$ orbitals in the basis space.  An even number of $L=2$ orbitals produce a larger 
$Q_{20}$ result with the bare operator in the truncated basis.  This signals that the mixing generated by the gs eigenvector in the truncated basis has favorable phases for contributions to $Q_{20}$ with an even number of $L=2$ orbitals.

On the other hand, the magnetic dipole operator shown in Fig. \ref{Summary_OLS_effops} reflects minimal renormalization effects. Note that the scale for these results is only a couple of percent in the smallest model spaces.  This is consistent with a number of many-body applications that, with increasing model spaces, show the magnetic moments are well converged in contrast to other long-range observables such as $r_{RMS}$, $Q_{20}$ and $B(E2)$ operators.

For each observable in Fig. \ref{Summary_OLS_effops}, the size of the effects in smaller model spaces may, at first, appear large compared with the systematic study conducted in Ref.~\cite{Stetcu:2004wh} showing long-range operators receive only minor renormalization effects from the OLS procedure.  However, it is important to note that our test two-nucleon problem is special in that we can treat the OLS renormalization exactly for all observables in all model spaces.  This contrasts the cases studied in Ref.~\cite{Stetcu:2004wh} where the OLS renormalization was performed at the two-nucleon level but then applied in many-nucleon systems so that the induced many-nucleon correlation contributions to the effective electroweak operators were neglected.  Thus, as was emphasized in Ref. \cite{Stetcu:2004wh}, one must be cautious when drawing conclusions from applications in many-body applications using OLS renormalization limited to the two-nucleon level. 

\section{$A=48$ in the NCSM with a core approach}

\begin{figure}[!t]
\centerline{\includegraphics[width=0.94\textwidth]{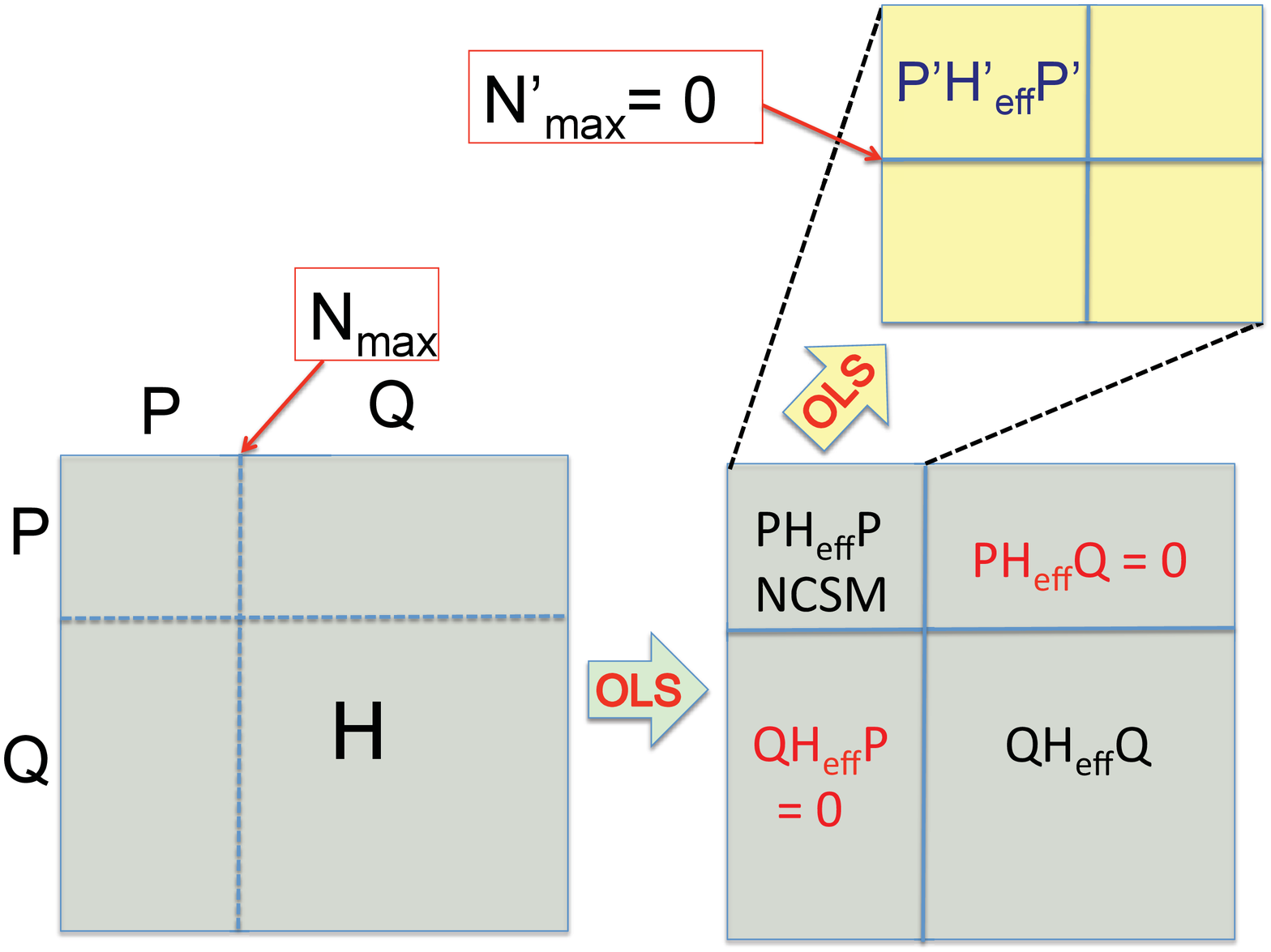}}
\caption{Schematic of the ``double OLS'' procedure that first takes results from a NCSM calculation
for a core system, using an OLS procedure for a model space defined by $N_{\rm max}$, 
as input to generate an effective interaction among valence nucleons in the $N_{\rm max}'=0$ model space 
as described in  Refs. \cite{Lisetskiy:2008ja,Dikmen:2015tla}. That is, the OLS procedure is first applied 
to derive a NCSM effective interaction for the full $A$-nucleon system resulting 
in the ``primary'' effective Hamiltonian $PH_{\rm eff}P$ for the chosen no-core basis space 
(the ``$P$-space'') indicated on the large square on the right of the figure in its upper left corner.  
The OLS procedure is applied again by using the NCSM results to derive the 
``secondary'' effective Hamiltonian $P'H'_{\rm eff}P'$ for the valence space 
(the $P'$-space with the smaller many-body cutoff $N'_{\rm max}$) indicated on the square 
in the upper right of the figure.}
\label{diagram}      
\end{figure}

There is considerable interest in pushing {\it ab initio} nuclear structure and nuclear reaction methods to heavier nuclei
and a number of approaches are under development. For the NCSM, the path forward has been defined in a series
of efforts \cite{Lisetskiy:2008ja, Dikmen:2015tla}.  Schematically, the approach adopts the NCSM for a chosen core such as $^{16}$O or $^{40}$Ca in as large a basis as feasible and uses the OLS renormalization for that basis.  An alternative would be to use the Similarity Renormalization Group (SRG) method~\cite{Glazek:1993rc,Wegner:1994,Bogner:2007rx,Hergert:2007wp}
for the NCSM treatment of the adopted $A_c$  nucleon ``core'' system.  In like manner, one solves the $A_c+1$ 
 nucleon and $A_c+2$  nucleon systems to obtain the eigenvalues and eigenfunctions. With the resulting eigenvalues and eigenfunctions, one then performs another OLS treatment for the $N_{\rm max}=0$ 
 space or valence space
with 2 nucleons beyond the core to derive an effective 2-valence nucleon interaction.  This valence-only interaction is guaranteed to generate the same results in the $N_{\rm max}=0$ space as the original NCSM calculation for the $A_c+2$ nucleon system as demonstrated in Refs. \cite{Lisetskiy:2008ja,Dikmen:2015tla}. This logic is straightforwardly extended to derive a 3-valence nucleon effective interaction or even 4-valence nucleon effective interaction. This process is illustrated schematically in Fig. \ref{diagram}.

For an application in the pf-shell, such as the $A=48$ nuclei, we envision solving for $^{40}$Ca (the core), 
$^{41}$Ca (core + 1) and $^{42}$Ca (core + 2) systems in the NCSM with $N_{\rm max}=4$.  Later, we would
include the $^{42}$Ca (core + 3) system to obtain a valence 3-neutron effective interaction.  Once the second
OLS transformation is performed we would have the valence single-particle-energies and valence effective two-body interactions suitable for a standard shell model calculation of $^{48}$Ca.  Next, we would seek to confirm that this provides a reasonable
description of the properties of $^{48}$Ca.  Following that, we would proceed with additional calculations needed to evaluate the 
double beta-decays of $^{48}$Ca, both with neutrinos and without neutrinos.  Such studies will be valuable for benchmarking other nuclear structure approaches that are currently in use for evaluating nuclear matrix elements for double-beta decays in heavier nuclei.

Let us examine a few more of the specifics of the double OLS approach to the $A=48$ nuclei with a particular selection of ingredients.  Let us select an $NN + NNN$ interaction case for the NCSM treatment of $^{40}$Ca and the $A=41$ and $A=42$ nuclei in the $N_{\rm max}=4$ space with OLS renormalization.  The largest matrix encountered is that of $^{42}$Sc with $M$-scheme dimension $1,211,160,184$ and $54 \times 10^{12}$ nonzero many-body matrix elements.  We would need to converge a minimum of 60 eigenvalues and eigenvectors to perform the second OLS transformation needed for the $195$ valence $NN$ interaction matrix elements with good $J,\ T$. The reason for the minimum of 60 is that we should obtain those eigenvalues whose eigenvectors have significant overlap with the pf-space and have the requisite number for each $J,\ T$ combination. These calculations seem likely to be feasible with current technologies.

\section{Future prospects}

Most of our applications have focused on light nuclei with atomic number $A \leq 16$ where our
theoretical many-body methods have achieved successes with Leadership-Class facilities.
However, the frontiers of our field include applications to heavier nuclei and 
utilizing new and improved interactions from chiral EFT.  At the same time, we aim to evaluate 
observables with increasing sophistication using their operators also derived within
chiral effective field theory.  We sketched a near-term project for the $A=48$ nuclei. Our approach, which
aims to make contact with experimental and other theoretical efforts in double-beta decay, 
is but one exciting example of frontier research with {\it ab initio} nuclear theory.  
Others are also addressed at this same meeting.

We continue to face the dual challenge of advancing the underlying 
theoretical physics at the same time as advancing the algorithms to keep pace with the growth
in the size and complexity of Leadership-Class computers.  Recent history of these
efforts, with the substantial support of the funding agencies,
indicates we are experiencing a ``Double Moore's Law" rate of improvement~--- i.\;e.
Moore's Law for hardware improvements and a simultaneous Moore's Law improvement in
the algorithms/software.  We value this continued support of the funding agencies which 
has been and continues to be critical for our 
multi-disciplinary collaborations as well as their support of the growth in Leadership-Class facilities.  
This continued support will allow us to achieve the full discovery potential of computational 
fundamental physics.

\section{Acknowledgements}

This work was supported in part by the US Department of
Energy (DOE) under Grant Nos.~DE-FG02-87ER40371, DESC0008485
(SciDAC-3/NUCLEI), DE-SC0018223 \\ (SciDAC-4/NUCLEI) and DE-SC0015376 (DOE Topical 
Collaboration in Nuclear Theory for Double-Beta Decay and Fundamental Symmetries).
A portion of the computational resources were
provided by the National Energy Research Scientific Computing Center
(NERSC), which is supported by the US DOE Office of Science, and by 
an INCITE award, ``Nuclear Structure and Nuclear Reactions'', from the US DOE
Office of Advanced Scientific Computing.  This research also used 
resources of the Argonne Leadership Computing Facility, 
which is a DOE Office of Science User Facility supported 
under Contract DE-AC02-06CH11357.

\end{document}